\documentclass[final,authoryear,5p,times,twocolumn]{elsarticle}
\usepackage{amsmath}
\usepackage{amssymb} 

\usepackage{tikz}
\usetikzlibrary{arrows,matrix,positioning,patterns}
\usetikzlibrary{calc}
\usepackage{pgfplots} 
\usetikzlibrary{external}
\usetikzlibrary{plotmarks}

\newdefinition{rmk}{Remark}
\newproof{pf}{Proof}
\newproof{pot}{Proof of Theorem \ref{thm2}}

\usepackage[ruled, vlined, linesnumbered]{algorithm2e}

\usepackage{amsthm}

\usepackage{amsmath}
\usepackage{amssymb} 

\usepackage{eurosym}
\usepackage{colortbl}

\usetikzlibrary{arrows.meta,calc,decorations.markings,math,arrows.meta}

\usepackage[labelformat=simple]{subcaption}
\DeclareCaptionLabelSeparator{periodspace}{.\quad}
\usepackage{color,framed}
\usepackage[utf8]{inputenc} 
\usepackage{tabularx}

\usepackage{rotating}
\usepackage{graphicx}
\usepackage{lscape}
\allowdisplaybreaks 
\usepackage{longtable}

\usepackage[utf8]{inputenc}
\usepackage{pgfplots}
\usepgfplotslibrary{groupplots} 
\pgfplotsset{compat=1.3}

\makeatletter
\def\ps@pprintTitle{%
 \let\@oddhead\@empty
 \let\@evenhead\@empty
 \def\@oddfoot{}%
 \let\@evenfoot\@oddfoot}
\makeatother

\begin{document}

\begin{frontmatter}
%
%
\title{GPU-accelerated Logistics Optimisation for Biomass Production\\ with Multiple Simultaneous Harvesters Tours, Fields and Plants}
\author{Mogens Graf Plessen\corref{cor1}}
\cortext[cor1]{\texttt{mgplessen@gmail.com}}

%
%
%
%

\begin{abstract}
Within the context of biomass production, this paper proposes a method for logistics optimisation. Starting from a headquarter multiple tours are to be executed simultaneously by groups of harvesting units (HUs) and support units (SUs) to first harvest biomass from multiple agricultural fields, before supplying the biomass to multiple biogas plants (BPs) via shuttling SUs. This problem is relevant on a larger scale in particular for contractors. This problem is complex since there are three interconnected optimisation levels: (i) the assignment of BPs to tours and the ordering of BPs assigned to each tour, (ii) the assignment of fields to BPs and the ordering of fields assigned to each BP, and (iii) determining the number of HUs and SUs assigned to each tour, whereby different HUs and SUs may in general have different working rates and loading capacities. Problem modeling and a solution method are discussed. For the latter, a GPU-accelerated heuristic search algorithm is proposed. For the former, an optimisation criterion minimising both total accumulated path length and the maximum completion time over all harvesters tours, an embedded local minimisation for the assignment of SUs to tours, and demand fulfilment constraints for BPs are discussed. In stochastic simulation experiments it is found that permitting unconstrained assignment of any available field that a contractor services to any available BP, independent of their ownerships, is crucial (i) to attain maximum path length savings, and (ii) also to best balance and minimise uniform completion times over all harvesters tours such that weather-dependent harvesting windows can be exploited optimally.
\end{abstract} 
\end{frontmatter}


\section{Introduction\label{sec_intro}}

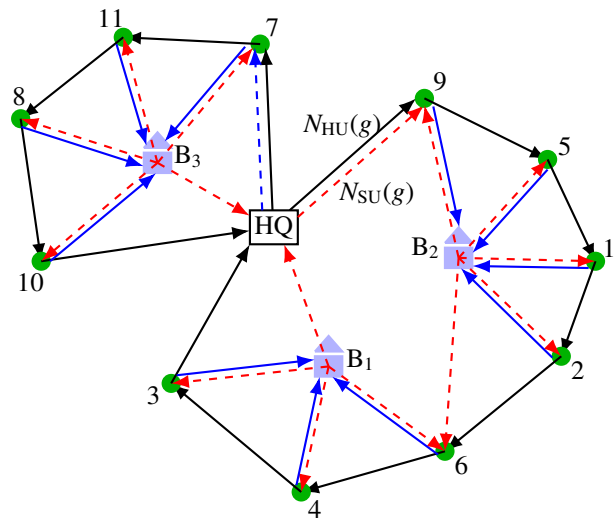
\begin{figure}
\centering
\vspace{0.cm}
\begin{tikzpicture}[thick,scale=0.9, every node/.style={scale=1.0}]
%
\draw[black,fill=white,fill opacity=0.2] (-0.35,-0.25) rectangle (0.35,0.25);
\node[color=black] (a) at (0,0) {HQ};
\draw[black,-{Latex[scale=1.0]}] (0.23, 0.25) -- (2.08,1.9); 
\node[color=black] (a) at (1,1.5) {$N_\text{HU}(g)$};
\draw[red,dashed,-{Latex[scale=1.0]}] (0.35, 0.15) -- (2.2,1.8); 
\node[color=black] (a) at (1.5,0.5) {$N_\text{SU}(g)$};
%
\draw[blue!30,fill=blue!30,fill opacity=1.0] (2.5,-0.6) rectangle (2.9,-0.3);
\draw[blue!30,fill=blue!30,fill opacity=1.0] (2.5,-0.23) -- (2.9,-0.23) -- (2.7,0) -- (2.5,-0.23);
\node[color=black] (a) at (2.22,-0.3) {B$_2$};
%
\fill[color=green!70!black] (2.2,1.9) circle [radius=4pt];
\node[color=black] (a) at (2.43,2.15) {9};
\draw[black,-{Latex[scale=1.0]}] (2.2,1.9) -- (4,1);
\draw[red,dashed,-{Latex[scale=1.0]}] (2.7,-0.45) -- (2.2,1.8); 
\draw[blue,-{Latex[scale=1.0]}] (2.33,1.78) -- (2.71,0); 
\fill[color=green!70!black] (4,1) circle [radius=4pt];
\node[color=black] (a) at (4.25,1.2) {5};
\draw[black,-{Latex[scale=1.0]}] (4,1) -- (4.7,-0.5);
\draw[red,dashed,-{Latex[scale=1.0]}] (2.7,-0.45) -- (4,1); 
\draw[blue,-{Latex[scale=1.0]}] (4,0.85) -- (2.9,-0.38); 
%
\fill[color=green!70!black] (4.7,-0.5) circle [radius=4pt];
\node[color=black] (a) at (4.9,-0.3) {1};
\draw[black,-{Latex[scale=1.0]}] (4.7,-0.5) -- (4.2,-1.9);
\draw[red,dashed,-{Latex[scale=1.0]}] (2.7,-0.45) -- (4.7,-0.5);
\draw[blue,-{Latex[scale=1.0]}] (4.6,-0.6) -- (2.9,-0.57); 
%
\fill[color=green!70!black] (4.2,-1.9) circle [radius=4pt];
\node[color=black] (a) at (4.44,-2.1) {2};
\draw[black,-{Latex[scale=1.0]}] (4.2,-1.9) -- (2.5,-3.3);
\draw[red,dashed,-{Latex[scale=1.0]}] (2.7,-0.45) -- (4.2,-1.9);
\draw[blue,-{Latex[scale=1.0]}] (4.07,-1.92) -- (2.73,-0.6); 
%
%
\draw[blue!30,fill=blue!30,fill opacity=1.0] (0.6,-2.2) rectangle (1,-1.9);
\draw[blue!30,fill=blue!30,fill opacity=1.0] (0.6,-1.83) -- (1,-1.83) -- (0.8,-1.6) -- (0.6,-1.83);
\draw[red,dashed,-{Latex[scale=1.0]}] (0.8,-2.05) -- (0.15,-0.25);
\node[color=black] (a) at (1.27,-1.95) {B$_1$};
%
\fill[color=green!70!black] (2.5,-3.3) circle [radius=4pt];
\node[color=black] (a) at (2.74,-3.5) {6};
\draw[black,-{Latex[scale=1.0]}] (2.5,-3.3) -- (0.4,-3.9);
\draw[red,dashed,-{Latex[scale=1.0]}] (2.7,-0.45) -- (2.5,-3.3);
\draw[red,dashed,-{Latex[scale=1.0]}] (0.8,-2.05) -- (2.5,-3.3);
\draw[blue,-{Latex[scale=1.0]}] (2.4,-3.35) -- (0.85,-2.2);
%
\fill[color=green!70!black] (0.4,-3.9) circle [radius=4pt];
\node[color=black] (a) at (0.6,-4.15) {4};
\draw[black,-{Latex[scale=1.0]}] (0.4,-3.9) -- (-1.5,-2.3);
\draw[red,dashed,-{Latex[scale=1.0]}] (0.8,-2.05) -- (0.4,-3.9);
\draw[blue,-{Latex[scale=1.0]}] (0.33,-3.8) -- (0.67,-2.2);
%
\fill[color=green!70!black] (-1.5,-2.3) circle [radius=4pt];
\node[color=black] (a) at (-1.76,-2.48) {3};
\draw[black,-{Latex[scale=1.0]}] (-1.5,-2.3) -- (-0.35,-0.25);
\draw[red,dashed,-{Latex[scale=1.0]}] (0.8,-2.05) -- (-1.5,-2.3);
\draw[blue,-{Latex[scale=1.0]}] (-1.42,-2.18) -- (0.6,-1.95);
%
%
\draw[blue!30,fill=blue!30,fill opacity=1.0] (-1.9,0.8) rectangle (-1.5,1.1);
\draw[blue!30,fill=blue!30,fill opacity=1.0] (-1.9,1.17) -- (-1.5,1.17) -- (-1.7,1.4) -- (-1.9,1.17);
\node[color=black] (a) at (-1.25,1.05) {B$_3$};
\draw[red,dashed,-{Latex[scale=1.0]}] (-1.7,0.95) -- (-0.35,0.17);
%
\fill[color=green!70!black] (-0.2,2.7) circle [radius=4pt];
\node[color=black] (a) at (-0.02,2.95) {7};
\draw[black,-{Latex[scale=1.0]}] (-0.02,0.25) -- (-0.12,2.65);
\draw[blue, dashed,-{Latex[scale=1.0]}] (-0.17,0.25) -- (-0.28,2.65);
\draw[black,-{Latex[scale=1.0]}] (-0.2,2.7) -- (-2.2,2.8);
\draw[red,dashed,-{Latex[scale=1.0]}] (-1.7,0.95) -- (-0.3,2.65);
\draw[blue,-{Latex[scale=1.0]}] (-0.42,2.66) -- (-1.6,1.26);
%
\fill[color=green!70!black] (-2.2,2.8) circle [radius=4pt];
\node[color=black] (a) at (-2.38,3.08) {11};
\draw[black,-{Latex[scale=1.0]}] (-2.2,2.8) -- (-3.7,1.6);
\draw[red,dashed,-{Latex[scale=1.0]}] (-1.7,0.95) -- (-2.2,2.8);
\draw[blue,-{Latex[scale=1.0]}] (-2.3,2.7) -- (-1.86,1.2);
%
\fill[color=green!70!black] (-3.7,1.6) circle [radius=4pt];
\node[color=black] (a) at (-3.72,1.93) {8};
\draw[black,-{Latex[scale=1.0]}] (-3.7,1.6) -- (-3.4,-0.5);
\draw[red,dashed,-{Latex[scale=1.0]}] (-1.7,0.95) -- (-3.7,1.6);
\draw[blue,-{Latex[scale=1.0]}] (-3.69,1.48) -- (-1.9,0.9);
%
\fill[color=green!70!black] (-3.4,-0.5) circle [radius=4pt];
\node[color=black] (a) at (-3.56,-0.81) {10};
\draw[black,-{Latex[scale=1.0]}] (-3.4,-0.5) -- (-0.35,-0.05);
\draw[red,dashed,-{Latex[scale=1.0]}] (-1.7,0.95) -- (-3.4,-0.5);
\draw[blue,-{Latex[scale=1.0]}] (-3.27,-0.5) -- (-1.72,0.8);
%
%
%
\end{tikzpicture}
\caption{Problem sketch with exemplary $N_T=2$ harvesters tours (paths which start and end at the same graph vertex, the headquarter), $N_B=3$ biogas plants (blue houses) and $N_F=11$ fields (numbered green dots indicate centroids of fields). Paths traveled by harvesting units (HUs), empty support units (eSUs) and loaded support units (lSUs) are denoted by black, dashed-red and blue arrows, respectively. All mobile agents start at a central headquarter (HQ), whereby the total number of HUs and SUs is limited, i.e., $N_\text{HU}^\text{total}$ and $N_\text{SU}^\text{total}$. All fields (numbered for identification) provide an expected supply of biomass, whereas all BPs have a minimum demand. Subject to constraints the optimisation tasks are: (i) assignments of BPs to tours, (ii) their ordering for each assignment, (iii) assignment of fields to BPs, (iv) their ordering for each assignment, (v) assignment of $N_\text{HU}(g),\forall g=1,\dots,N_T$, and (vi) assignment of $N_\text{SU}(g),\forall g=1,\dots,N_T$.}
\label{Fig_problemVisualisation}
\end{figure}

Biomass belongs to the trending class of renewable energy sources. The two main benefits are (i) its versatility in generating not only electricity but also heat and biofuels, and (ii) its storage capability to generate energy on-demand. The typical supply chain for agricultural biomass production involves planting, harvesting, road transport, storage and utilisation for energy exploitation. A critical characteristic is that typically only a very limited and weather-dependent optimal harvesting period is available. This demands optimised logistical operations. The largest fraction of cost in biomass energy generation originates from logistics operations (\cite{rentizelas2009logistics}).

Optimising logistics operations typically demands to initially formulate a \emph{system model}. There exist many, each typically tailored to the specific application at hand. For the architecture of a biomass supply chain simulation model and functional modeling see \cite{pavlou2016simulation} and \cite{pavlou2016functional}. \cite{ebadian2011new} presented a biomass logistics model for demand fulfilment and storage management. \cite{sokhansanj2006development} proposed a model to simulate the flow of biomass from a field to a biorefinery involving collection, storage, and transport operations. \cite{marufuzzaman2017managing} discussed a dynamic multi-modal facility location model to alleviate the impacts of congestion on biomass supply chain performance due to seasonality of biomass. They proposed a mixed integer nonlinear program to trade-off investment, transportation, and congestion management decisions. A linear approximation of the model was then solved using a hybrid Benders-based rolling horizon algorithm. In previous work, \cite{marufuzzaman2016designing} presented a dynamic multi-modal transportation network model for the delivery of biomass to biofuel plants subject to biomass supply fluctuations and hedging against natural disasters. \cite{cundiff1997linear} presented a linear programming-based biomass delivery model. General opportunities and challenges in biomass and biofuel supply chain modeling were summarised by \cite{yue2016biomass}. A review for the design of biomass feedstock supply chain models was provided by \cite{sun2018biomass}. Other reviews are by \cite{malladi2018biomass}, \cite{atashbar2016modeling}, \cite{castillo2014metaheuristic}, \cite{yue2014biomass} and \cite{gold2011supply}.

\begin{table}
\centering
\begin{tabular}{|ll|}
\hline
\multicolumn{2}{|c|}{MAIN NOMENCLATURE}\\
\multicolumn{2}{|l|}{Symbols}\\
$\mathcal{B}$ & Set of all biogas plants (BPs).\\
$\mathcal{F}$ & Set of all fields.\\
$\mathcal{F}_b$ & Set of fields assigned to a BP $b\in\mathcal{B}$.\\
$\mathcal{N}_\text{SU}^\text{small}$ & List of $N_\text{SU}^\text{small}(g),~\forall g=1,\dots,N_T$.\\
$\mathcal{N}_\text{SU}^\text{large}$ & List of $N_\text{SU}^\text{large}(g),~\forall g=1,\dots,N_T$.\\
$\mathcal{T}_\text{HU}^\text{compl}$ & List of completion times for all harvesters tours.\\
$\mathcal{T}_\text{HU}^\text{wait}$ & List of average waiting times for all tours.\\
$\mathcal{C}$ & List of costs for all tours, $g=1,\dots,N_T$, (km).\\
$C$ & Total accumulated cost, (km).\\
$\Delta C^\star$ & Cost difference w.r.t. a baseline, (km) or ($\%$).\\
$N_\text{iters}$ & Hyperparameter in Algorithm \ref{alg_1}.\\
$N_{T}$ & Number of active tours.\\
$N_{B}$ & Number of BPs.\\
$N_{F}$ & Number of fields.\\
$N_\text{HU}^\text{total}$ & Total number of HUs.\\
$N_\text{SU}^\text{total}$ & Total number of SUs.\\
$N_\text{HU}(g)$ & Number of HUs assigned to tour $\{g\}_1^{N_T}$.\\
$N_\text{SU}(g)$ & Number of SUs assigned to tour $\{g\}_1^{N_T}$.\\
$N_{\text{SU},f}^\text{shuttle}$ & Number of shuttling drives from field $f\in\mathcal{F}$.\\
$T^\text{solve}$ & Solve time for running Algorithm \ref{alg_1}, (s).\\ 
$c_\text{biom}^\text{conv}$ & Lumped conversion parameter, (t~ha$^{-1}$).\\
$d_b^\text{min}$ & Minimum demand of biomass at BP $b\in\mathcal{B}$, (t).\\
$\epsilon$ & Hyperparameter in Algorithm \ref{alg_1}.\\
$h_f$ & Field size for $f\in\mathcal{F}$, (ha).\\
$l_\text{SU}^\text{load,small}$ & Loading capacity of a small SU, (t).\\
$l_\text{SU}^\text{load,large}$ & Loading capacity of a large SU, (t).\\
$\hat{s}_f$ & Expected biomass supply from field $f\in\mathcal{F}$, (t).\\
$t_\text{SU}^{\text{load},\text{small}}$ & In-field fill time required by small SUs, (min).\\
$t_\text{SU}^{\text{load},\text{large}}$ & In-field fill time required by large SUs, (min).\\
$v_\text{HU}^\text{area}$ & In-field working rate of HUs, (ha~h$^{-1}$).\\
$v_\text{HU}^\text{edge}$ & Road travel velocities of HUs, (km~h$^{-1}$).\\
$v_\text{SU}^\text{edge}$ & Road travel velocities of SUs, (km~h$^{-1}$).\\[3pt]
\multicolumn{2}{|l|}{Abbreviations}\\
BPs & Biogas Plants. \\
HQ &  Headquarter (central depot). \\
HUs &  Harvesting Units (e.g, corn harvesters). \\
SUs &  Support Units (e.g., harvest truck with tractor). \\
\hline
\end{tabular}
\vspace{-0.4cm}
\end{table}
%

More applications of biomass logistic operations are discussed, now also focussing explicitly on problem dimensions to later contrast the contribution of this paper. In \cite{bochtis2013flow} the finding of a permutation schedule for a number of geographically dispersed fields where sequential biomass handling operations have to be carried out (baling, loading and transport) was formulated as a flow shop problem with sequence dependent setup times (\cite{zandieh2006immune}). They considered two machinery systems, a baling system (tractor-baler) and a loading system (fork-lift and transport truck), working in series on a total of 4 fields and starting from a central depot. A sensitivity analysis was also provided that demonstrated the importance of accurate and predictable parameters in the optimisation problem. It was found that solution quality can degrade quickly otherwise. Related to baling, \cite{zamar2017constrained} proposed for the bale collection problem a constrained k-means and nearest neighbour approach. This problem occurs after harvest and baling operations of a crop and consists of defining the sequence in which bales spread over a field are collected. A method for scheduling of machinery fleets in biomass multiple-field operations was proposed by \cite{orfanou2013scheduling}. For their case study they considered grass harvesting on a total of 5 fields with a maximum of 2 mowing, 2 raking and 2 baling machines. In \cite{busato2019optimisation}, a method to determine the optimal number of transport units in individual fields during harvest for silage production is proposed, minimising operational costs subject to time constraints. For their field trials they considered 4 different locations with between 2 to 5 fields each. \cite{gracia2014application} discussed an application of the vehicle routing problem to the biomass collection problem. A hybrid approach based on genetic algorithms and local search methods was presented to solve a real case study, where they considered 1 central storage location, 12 vehicle routes and 146 orchards. \cite{aguayo2017corn} discussed a corn-stover harvest scheduling problem that arises when a cellulosic ethanol plant contracts with farmers to harvest corn stover after the grain harvest has been completed. The plant contracts a fleet of harvesting crews, which must be assigned by the plant scheduler to harvest fields as they are called in by the farmers over the harvest season. The objective is to minimize the total cost incurred by the plant by determining the optimal number of balers required, the type of balers to harvest each field, the number of periods needed to complete the harvest of each field, and the routing of balers over the planning horizon. For their case study they considered 1 cellulosic ethanol plant and 85 fields. 

Within this context the motivation and contribution of this paper is to propose a method for large-scale logistics optimisation for biomass production. Starting from a headquarter multiple tours are to be executed simultaneously by groups of harvesting units (HUs) and support units (SUs) to first harvest biomass from multiple agricultural fields, before supplying the biomass to multiple biogas plants (BPs) via shuttling SUs. The problem is sketched in Fig. \ref{Fig_problemVisualisation}.  This logistics problem is relevant on a larger scale in particular for contractors. In practice, it is still often solved manually by human schedulers. The previous paragraph mentioned some problem dimensions from the literature to explicitly contrast the large-scale problems addressed in this study. These are inspired by a real-world scenario with 7 tours, 20 biogas plants, 1200 corn fields, 7 HUs and a total of 42 SUs operated on an area of 80km$\times$80km. In addition, part of this papers's contribution is to quantify optimisation-based gains with respect to the working practice. This is achieved by comparing and initialising the optimisation procedure with a solution that is replicating the in practice common decision process. This logistics problem is not the least complex since there are three interconnected optimisation levels: (i) the assignment of BPs to tours and the ordering of BPs assigned to each tour, (ii) the assignment of fields to BPs and the ordering of fields assigned to each BPs, and (iii) determining of the number of HUs and SUs, $N_\text{HU}(g)$ and $N_\text{SU}(g)$, assigned to each tour $g=1,\dots,N_T$, whereby different HUs and SUs may have different working rates and loading capacities. The dimensions of the problem considered in this paper are significantly larger than those of case studies reported in the literature. Likewise, to the author's knowledge, a solution based on GPU (graphics processing unit) computation has not yet been proposed within the agricultural biomass supply chain context. To summarise, there is a research gap with respect to (i) large-scale logistics optimisation for biomass production from the perspective of contractors, and (ii) presentation of a corresponding parallelisable solution approach suitable for GPU computation. 

The remaining paper is organised as follows. Materials and methods as well as numerical results are discussed in \S \ref{sec_problModeling} and \S \ref{sec_IllustrativeEx}, before concluding.

\section{Material and methods\label{sec_problModeling}}

The proposed system model, including constraints and cost function, is explained by Fig. \ref{Fig_problemVisualisation}, \S \ref{subsec_problemMdl_cstrts} and \S \ref{subsec_distBasedCostFcn}-\ref{subsec_problemMdl_alternativeCost}. The proposed optimisation algorith is presented in \S \ref{subsec_mainAlg}. Variations are briefly discussed in \S \ref{subsec_problSoln_variations}. Proposed solution method and all simulation experiments from \S \ref{sec_IllustrativeEx} were implemented from scratch in Cuda C++ and run on an Intel i7-7700K CPU{@}4.20GHz$\times$8 and a TitanV-GPU.

\begin{table}
\centering
\vspace{0.5cm}
\def\arraystretch{1.15}
\begin{tabular}{lll}
\hline
\rowcolor[gray]{0.9} 
Symbol  & Unit & Value\\\hline        
$N_\text{HU}^\text{total}$   & - &  7 \\
$N_\text{SU}^\text{small}$   & - &  14 \\
$N_\text{SU}^\text{large}$   & - &  28 \\
$l_\text{SU}^\text{load,small}$   & t & 12.5 \\ 
$l_\text{SU}^\text{load,large}$   & t & 16.5\\ 
$t_\text{SU}^\text{load,small}$ &  min & 6 \\
$t_\text{SU}^\text{load,large}$ &  min & 8 \\ 
$v_\text{SU}^\text{edge}$    & km~h$^{-1}$ & 40\\
$v_\text{HU}^\text{edge}$    & km~h$^{-1}$ & 40 \\ 
$v_\text{HU}^\text{area}$    & ha~h$^{-1}$ & 2.5 \\
$c_\text{biom}^\text{conv}$  & t~ha$^{-1}$ & 40  \\ 
\hline
\end{tabular}
\caption{Numerical values of all system parameters are summarised as empirical averages from real-world operations in Northern Germany. All remaining location-dependent parameters are: (i) $\hat{s}_f,\forall f\in\mathcal{F}$, (ii) $d_b^\text{min},~\forall b\in\mathcal{B}$, (iii) $h_f,~\forall f\in\mathcal{F}$, and (iv) one transition matrix indicating the distance costs for a path network connecting HQ, all BPs and all fields.}
\label{tab_problemData}
\end{table}

\subsection{Constraints and parameters for problem modeling\label{subsec_problemMdl_cstrts}}

First, there are multiple groups of HUs and SUs dispatched to multiple BPs and fields. Since typically $N_B>N_T$, multiple BPs must be served along each tour. The number of active tours is always bounded by the number of total available HUs, i.e., $N_T\leq N_\text{HU}^\text{total}$. However, there must not necessarily be $N_\text{HU}^\text{total}$ active tours since multiple HUs can in general operate along the same tours (e.g., to increase working rates). HUs never drive to BPs. Instead, they travel from field to field. Thus, it is assumed that they are maintained and refilled with fuel directly on the fields without returning to HQ. Similarly, also SUs return to HQ only after harvest completion on all assigned fields.

Second, it holds that
\begin{equation}
N_\text{HU}^\text{total} = \sum_{g=1}^{N_T} N_\text{HU}(g), \quad N_\text{SU}^\text{total} = \sum_{g=1}^{N_T} N_\text{SU}(g),
\end{equation}
and, in general, $N_\text{HU}(g)=\sum_{l=1}^{N_\text{HU}^\text{classes}} N_\text{HU}^l(g)$ and $N_\text{SU}(g)=\sum_{l=1}^{N_\text{SU}^\text{classes}} N_\text{SU}^l(g),~\forall g=1,\dots,N_T$, where $N_\text{HU}^l(g),\forall l=1,\dots,N_\text{HU}^\text{classes}$ and  $N_\text{SU}^l(g),\forall l=1,\dots,N_\text{SU}^\text{classes}$ denote different classes of HUs and SUs that may differ by working rates and loading capacities, respectively. 

Third, all BPs must be served a minimum biomass supply, 
\begin{equation}
\sum_{f\in\mathcal{F}_b} \hat{s}_f \geq d_b^\text{min},~\forall b\in\mathcal{B},
\label{eq_minDemand_cstrts}
\end{equation}
where $\mathcal{B}$, $\mathcal{F}_b\subset \mathcal{F}$, $d_b^\text{min}$, and $\hat{s}_f$ denote the set of BPs, the set of fields assigned to a specific BP $b\in\mathcal{B}$, the minimum demand of biomass, and the expected supply from each field $f\in\mathcal{F}$, respectively. It is assumed that there is no upper bound on the maximum demand for each BP. Thus, good storage capabilities at BPs are assumed (\cite{rentizelas2009logistics}).

Fourth, each field supplies a certain expected supply that must  be assumed at the time of logistics planning. For simplicity, a linear relationship is assumed for the mapping from field size to biomass, $\hat{s}_f = \hat{c}_f h_f ,\forall f\in\mathcal{F}$, where field size and a lumped parameter are denoted by $h_f$ and $\hat{c}_f$, respectively. For simulations, a constant, $c_\text{biom}^\text{conv}=\hat{c}_f,\forall f\in\mathcal{F}$, is assumed and thus,
\begin{equation}
\hat{s}_f = c_\text{biom}^\text{conv} h_f ,\forall f\in\mathcal{F}.
\end{equation}   

Fifth, SUs are assumed to have a loading capacity limit denoted by $l_\text{SU}^{\text{load},l}$. For any given field $f\in\mathcal{F}_b$ the number of shuttling drives to its assigned BP is thus 
\begin{equation}
N_{\text{SU},f}^{\text{shuttle},l} =\left\lceil \hat{s}_f / l_\text{SU}^{\text{load},l}\right\rceil,~l=1,\dots,N_\text{SU}^\text{classes},\label{eq_NshuttlingDrives}
\end{equation}
where $\lceil \cdot \rceil$ denotes the ceiling operator.

Sixth, a comment about the path network is made. In case locations are not directly adjacent, shortest paths are implicitly computed along the path network to connect two locations.

Seventh, remaining relevant parameters are (i) average travel velocities along the path network for HUs and SUs, $v_\text{HU}^\text{edge}$ and $v_\text{SU}^\text{edge}$, which are assumed to be uniform over all mobile machinery, (ii) the in-field area working rates (ha/h) of HUs, $v_\text{HU}^{\text{area},l},~\forall l=1,\dots,N_\text{HU}^\text{classes}$, and (iii) fill times required to load up SUs, $t_\text{SU}^{\text{load},l},~\forall l=1,\dots,N_\text{SU}^\text{classes}$, which are assumed to be proportional to loading capacities, $l_\text{SU}^{\text{load},l}$. These parameters are assumed to be uniform among the path network and all fields. 

For clarity, a compact summary of final numerical values of all system parameters is provided in Table \ref{tab_problemData}.

\subsection{Objective function: Path length-based cost metric\label{subsec_distBasedCostFcn}}

The cost metric used for the evaluation of different solution candidates is selected as \emph{path length}-based. Thus, cost is measured as the total distance accumulated by all HUs and SUs involved for mission completion. Cost is abbreviated by symbol $C$.  An extending alternative \emph{monetary units}-based cost function, that was considered, is discussed further below. 
 
A path length-based cost metric has several advantages. There are comparatively few paramaters that need to be identified from data. Furthermore, non-negativity of path lengths implies existence of a global optimal solution. Ultimately, all optimisation tasks according to Fig. \ref{Fig_problemVisualisation} are addressed since accumulated total distance is clearly affected by (i) assignment of BPs to tours, (ii) their ordering within each tour, (iii) assignment of fields to BPs, (iv) their ordering for each assignment, (v) assignment of $N_\text{HU}(g),\forall g=1,\dots,N_T$, and (vi) assignment of $N_\text{SU}(g),\forall g=1,\dots,N_T$. 
 
The cost function is evaluated tracing traveled distances according to Fig. \ref{Fig_problemVisualisation}. Two aspects are emphasised. First, the assigned number of HUs and SUs, $N_\text{HU}(g)$ and $N_\text{SU}(g)$, multiply distances for each tour. Second, the number of shuttling drives according to \eqref{eq_NshuttlingDrives}, $N_{\text{SU},f}^{\text{shuttle},j}$, is multiplied by 2 in order to account for back-and-forth distances for SUs that shuttle between fields and their assigned BPs.

\subsection{Extension: Alternative cost function\label{subsec_problemMdl_alternativeCost}}

Above cost function is motivated (i) from the perspective of a contractor with full-time hired employees reimbursed at a fixed monthly rate, and (ii) with the objective of having to deal with only few and easy-to-quantify parameters. In a different setting with hourly reimbursements and more parameters an alternative \emph{monetary units}-based cost function may be derived.  By first translating distance to ``time'' as the dependent variable through knowledge of average travel speeds along all edges of the path network and in-field working rates of HUs, the two main cost positions can then be identified as: (i) hourly driver rates for all HUs and SUs, and (ii) hourly fuel consumption rates times Diesel prices for HUs and SUs. Here, complexity arises since fuel consumption varies for in-field work, along rural, gravel or paved roads, and dependent on whether SUs are loaded or empty during their shuttling services. Likewise, fuel prices may vary substantially. These variations and uncertainties in parameters make monetary units-based cost function modeling more complex and less reliable with respect to practical employment.

Finally, note that all cost functions derived from translating distance to time as the dependent variable before multiplications by different hourly cost rates can always be reformulated as a nonlinearly weighted version of the distance-based cost function from \S \ref{subsec_distBasedCostFcn}. Constraints and likewise the solution algorithm are not affected by neither cost function choice.

\subsection{Decision variables\label{subsec_decisionVariables}}
 
The optimisation tasks are: (i) assignments of BPs to tours, (ii) their ordering for each assignment, (iii) assignment of fields to BPs, (iv) their ordering for each assignment, (v) assignment of $N_\text{HU}(g),\forall g=1,\dots,N_T$, and (vi) assignment of $N_\text{SU}(g),\forall g=1,\dots,N_T$. For final experimental results in \S \ref{sec_IllustrativeEx}, four different setups are further distinguished that differ by fixing specific assignments, e.g., the assignment of specific fields to BPs. Thereby, working practice is emulated, where the contractor must frequently respect contractual agreeements and ownerships of different BP operators,
field owners and leasing parties.

\begin{algorithm}
\DontPrintSemicolon
Assign BPs to tours and order their visit within each tour.\;
Assign fields to BPs.\;
Order the visit of fields assigned to each BP.\;
Summarise the solution of Steps 1-3 as $x^\star$, compute the resulting cost $C^\star$, and set $x_\text{old}\leftarrow x^\star$.\;
\For{$N_\text{iters}$}
{
Update $x$ based on $x_\text{old}$, $x^\star$ and the exploration heuristic of \eqref{eq_xUpd_explorHeuristic}.\;
\If{CPU (central processing unit)-version}
{
Randomly remove a vertex (field or BP) from $x$ and randomly reinsert that vertex into $x$ following insertion heuristics. Check constraints, compute cost if feasible, and update $x^\star$ and $C^\star$ if improving.\;
Set $x_\text{old}\leftarrow x$.\;
}\ElseIf{GPU (graphics processing unit)-version}
{
Broadcast $x$ to GPU-workers $i=1,\dots,n$, and set $x_i,~ x_i^\star \leftarrow x$.\;
\textbf{On GPU}: Conduct Step 8 for each GPU-worker, and compute cost $C_i$.\;
Return $C_i$ of each worker $i=1,\dots,n$ to the host.\;
Compute $i^\star = \arg\min_i~ \{C_i\}_1^n$. Reconstruct $x_{i^\star}$ on the host.\;
If $C_{i^\star} < C^\star$, update $x^\star \leftarrow x_{i^\star}$ and $C^\star \leftarrow C_{i^\star}$.\;
Set $x_\text{old}\leftarrow x_{i^\star}$.\;
}
}
%
\caption{Logistics for Biomass Production}
\label{alg_1}
\end{algorithm}

\subsection{Parallelisable optimisation algorithm \label{subsec_mainAlg}}

Algorithm \ref{alg_1} is proposed to solve the problem formulated in \S \ref{subsec_problemMdl_cstrts}-\ref{subsec_decisionVariables}. Several explanatory comments are made. Steps 1-4 represent the solution initialisation and replicate a deterministic sequential solution procedure a human scheduler may implicitly conduct. Step 1 involves (i) assigning an approximately equal number of HUs to each tour, e.g., $N_\text{HU}(g)=N_B/N_T,~\forall g=1,\dots,N_T-1$, with $N_\text{HU}(N_T)$ capturing the integer residual, and (ii) sequential \emph{nearest neighbour searches} from HQ to a closest first BP, from the first BP to a closest second BP, and so forth until all BPs are assigned to all tours and ordered.

Step 2 involves the assignment of fields to BPs. Guiding notion is that every BP has a minimum demand for biomass, $d_b^\text{min},~\forall b\in\mathcal{B}$. Therefore, fields are picked and assigned to their spatially closest BP that is not yet supplied up to its minimum demand. Once all BPs are supplied with their minimum demand, all remaining (not yet assigned) fields are assigned to their spatially closest BP. This is enabled by assumption according to problem modeling of \S \ref{sec_problModeling} about the absence of any upper bound on supplied biomass and the assumption about good storage capabilities at BPs.   

Step 2 was not yet concerned about the ordering of fields for each BP-assignment. This is conducted as part of Step 3 in form of sequential nearest neighbour searches from the HQ to the first field of its assigned BP, from that field to a closest second field assigned to the same BP, and so forth until fields assigned to each BP are ordered.

In Step 4, the solution of Steps 1-3 is summarised by generic vector $x^\star$ before its cost is computed. The initialisation solution is, in general, suboptimal because of its sequential derivation. Furthermore, it is fully deterministic if fields in Step 2 are selected according to a deterministic rule such as, e.g., their field identification numbers. In Step 2, stochasticity may be inserted by randomly drawing fields. In \S \ref{sec_IllustrativeEx} about numerical examples, the initialisation solution will be considered as the baseline reference in order to quantify any iterative solution improvement from conducting Steps 5-16 of Algorithm \ref{alg_1}.

In Step 6, $x$ is updated according to the rule,
\begin{equation}
x = \begin{cases} x^\star, & \text{if}~u<\epsilon,\\
                  x_\text{old}, & \text{otherwise}, \end{cases}
\label{eq_xUpd_explorHeuristic}
\end{equation}
where $u\in[0,1]$ is drawn as a uniform random variable and $\epsilon\in[0,1]$ is a hyperparameter. This update rule can be interpreted as trading off exploitation and exploration for the evolution of new $x$ candidates. This update rule is motivated by the presence of multiple optimisation layers including both the assignment of BPs to tours and fields to BPs. For example, suppose a BP is newly assigned to an alternative tour. Then, this assignment may initially be suboptimal because fields assigned to this BP are not yet optimally ordered to fit the new tour scenario. Thus, initially there is no progress with respect to $x^\star$. However, throughout the course of iterations and better orderings of fields, a new optimal $x^\star$ may eventually be found.

Step 8 involves a simple remove- and reinsert-operation for vertices, which can represent either a field or a BP. The insertion heuristic is discussed. First, in general any field can be removed from any BP-assignment and arbitrarily reassigned to any BP. Second and in contrast, in case the removed vertex is a BP it can be reassigned to any arbitrary tour or to a different ordering index within the existing tour. Importantly, in case a BP is reassigned, also  \emph{all} of the fields assigned to that BP are always also completely reassigned to the new tour or to the new ordering position, such that the attachment of fields to that BP is maintained. Notice that since in practice $N_F\gg N_B$, it is more likely to draw fields instead of BPs. Variations can be thought of. For example, in \S \ref{sec_IllustrativeEx} an in practice important variation is discussed where the assignment of fields to BPs is \emph{fixed} (e.g., due to different ownerships or contractual agreements) and only the sequence in which the fields are harvested can be optimised. Furthermore, the reassignment of BPs may ultimately cause a tour to become \emph{inactive}, i.e., if the last and only BP of a tour is reassigned more efficiently to an alternative tour. This is done on purpose since it is a priori not obvious what the most cost-efficient number of active tours is.

Throughout Algorithm \ref{alg_1} checking constraints implies to verify if for a given candidate $x$ the minimum biomass supply constraints in \eqref{eq_minDemand_cstrts} are satisfied for all BPs. If this is not the case the $x$ candidate is discarded (and its cost not evaluated).

If all constraints are satisfied the cost for a candidate solution $x$ is evaluated. This involves 3 steps as discussed next.

First, the assignment of HUs to tours is discussed. To obtain an intuition, it is here focused on the case of uniform HUs for $N_\text{HU}(g),~\forall g=1,\dots,N_T$. The assignment method for the case of non-uniform HUs with different working rates is mentioned later. The general strategy for the assignment of HUs is to achieve approximately uniform completion times, $\mathcal{T}_\text{HU}^\text{compl}(g)$, over all harvesters tours in order to simultaneously exploit optimal weather-dependent harvesting windows. Therefore, a specific constraint is enforced minimising the maximum completion time over all harvesters tours to rank GPU-candidates solutions. Given road travel velocities and in-field working rates as well as all distances and field sizes the completion times for all harvesters tours can be calculated. Initially, the average number of HUs is assigned to each group. Then, in an iterative procedure tours with minimum and maximum completion times, $g_\text{min}$ and $g_\text{max}$, are identified, before their assigned number of HUs, $N_\text{HU}(g_\text{min})$ and $N_\text{HU}(g_\text{max})$, is decreased and increased by one HU, respectively. This is repeated until the maximum completion time over all tours cannot be further reduced.

Second, a logic must be devised for the assignment of SUs to all tours $g=1,\dots,N_T$. Before stating the proposed logic a preparatory remark is derived. Suppose a set $\mathcal{L}$ of $N_\text{SU}^\text{classes}$ different SU classes. Suppose further that these classes only differ by their in-field fill times, $t_\text{SU}^{\text{load},l},~\forall l\in\mathcal{L}$, but not by their traveling speed along roads (edges) between any pair of fields and BPs (vertices). Then, two different waiting times for HUs can be defined as,
\begin{align}
t_{\text{HU},f_{b_g},b_g}^{\text{wait},l_g^\text{case}}(g) = t_{f_{b_g},b_g}^\text{edge}  &- (N_\text{SU}^{l_g^\text{case}}(g)-1)\frac{t_\text{SU}^{\text{load},l_g^\text{case}}}{\sum_j N_\text{HU}^j(g)} \notag\\
&-\sum_{l\in\mathcal{L}\backslash\{l_g^\text{case}\}}\frac{N_\text{SU}^l(g) t_\text{SU}^{\text{load},l}}{\sum_j N_\text{HU}^j(g)},\label{eq_def_tHUwaufbl}
\end{align}
for the two cases $l_g^\text{case}\in\{ l_g^\text{largest}, l_g^\text{smallest}\}$, and where $t_{f_{b_g},b_g}^\text{edge}$ denotes the total travel time from field $f_{b_g}\in\mathcal{F}_{b_g}\subset\mathcal{F}$ to its assigned BP $b_g\in\mathcal{B}_g\subset\mathcal{B}$ and back (which can be computed based on path network distances and travel velocities), and where $l_g^\text{largest}\in\mathcal{L}$ and $l_g^\text{smallest}\in\mathcal{L}$ denote the SU classes with largest and smallest fill times, and where  $g\in\{1,\dots,N_T\}$. Negative $t_{\text{HU},f_{b_g},b_g}^{\text{wait},l_g^\text{largest}}(g)<0$ (and likewise $t_{\text{HU},f_{b_g},b_g}^{\text{wait},l_g^\text{smallest}}(g)<0$) imply that there are sufficiently many SUs assigned to the tours such that HUs can operate continuously without any interruption or waiting time. An interruption would occur if at a given time during harvest there is no SU to which a HU can transfer the harvested biomass. This is the case if $t_{\text{HU},f_{b_g},b_g}^{\text{wait},l_g^\text{largest}}(g)>0$. Comparing $t_{\text{HU},f_{b_g},b_g}^{\text{wait},l_g^\text{largest}}(g)$ and $t_{\text{HU},f_{b_g},b_g}^{\text{wait},l_g^\text{smallest}}(g)$, it is found after simplifications that $t_\text{SU}^{\text{load},l_g^\text{largest}} >t_\text{SU}^{\text{load},l_g^\text{smallest}}$. This implies that $t_{\text{HU},f_{b_g},b_g}^{\text{wait},l_g^\text{largest}}(g)$ is more critical and an update rule must be devised based on that equation and SUs must be assigned based on ordering according to their fill times. Algorithm \ref{alg_2} is therefore proposed for the assignment of SUs to tours.
\begin{algorithm}
\DontPrintSemicolon
Compute average waiting times for each tour,
\begin{equation*}
\mathcal{T}_\text{HU}^\text{wait}(g) \leftarrow \frac{1}{|\mathcal{B}_g|} \sum_{b_g\in \mathcal{B}_g} \frac{1}{|\mathcal{F}_{b_g} |} \sum_{f_{b_g}\in \mathcal{F}_{b_g}} t_{\text{HU},f_{b_g},b_g}^{\text{wait},l_g^\text{largest}}(g),
\end{equation*}
for all $g=1,\dots,N_T$.\;
Compute indices,
\begin{align*}
g^\text{max} &\leftarrow \underset{g\in\{1,\dots,N_T\}}{\text{arg}\max} \mathcal{T}_\text{HU}^\text{wait}(g) ,\\
g^{\text{max},\text{2nd}} &\leftarrow \underset{g\in\{1,\dots,N_T\}\backslash \{g^\text{max}\}}{\text{arg}\max} \mathcal{T}_\text{HU}^\text{wait}(g),\\
g^\text{min} &\leftarrow \underset{g\in\{1,\dots,N_T\}\backslash\{g^\text{max},g^\text{max,2nd}\}}{\text{arg}\min} \mathcal{T}_\text{HU}^\text{wait}(g).
\end{align*}\;
\vspace{-0.4cm}
Determine $\mathcal{L}(g^\text{min})$ as the set of available SU classes for tour $g^\text{min}$ and sort it according to rising fill times .\;
\For{$l\in\mathcal{L}(g^\text{min})$}
{
Remove one SU candidate of class $l\in\mathcal{L}(g^\text{min})$.\;
Add that SU to  the group of $\mathcal{L}(g^\text{max})$.\;
Recompute $\mathcal{T}_\text{HU}^\text{wait}(g^\text{max})$.\;
\If{$\mathcal{T}_\text{HU}^\text{wait}(g^\text{max})<\mathcal{T}_\text{HU}^\text{wait}(g^{\text{max},\text{2nd}})$ \emph{or} $l$ represents the largest SU-class element of $\mathcal{L}(g^\text{min})$,}
{
Accept that SU-assignment.\;
Break.\;
}
}
Repeat Steps 1-10 until $\mathcal{T}_\text{HU}^\text{wait}(g^\text{max})$ cannot be further reduced.\;
%
\caption{Assignment of SUs to Different Tours}
\label{alg_2}
\end{algorithm}

In case HUs are also non-uniform, an equivalent method is employed for the minimisation of the maximum completion time over all harvesters tours, where the set of available HUs classes is now sorted according to rising in-field working rates.

Finally, the third step for the evaluation of $x$ after the optimised assignment of different HUs and SUs to all groups is discussed. The best solution of all GPU-workers is selected according to the distance-based cost function from \S \ref{subsec_distBasedCostFcn}, whereby all candidate solutions $x_i$ that do not decrease the best-so-far maximum completion time  over all $N_T$ harvesters tours are discareded by setting their cost $C_i=\infty$. This ensures that approximately uniform completion times, $\mathcal{T}_\text{HU}^\text{compl}(g)$, result for all tours. Note that, in contrast to completion times of all $N_T$ harvesters tours, controlling \emph{waiting times} $\mathcal{T}_\text{HU}^\text{wait}(g),~\forall g=1,\dots,N_T$ is much easier since there are typically (many) more SUs in comparison to HUs.

Based on an argument of symmetry it was also considered to repeat Step 8 of Algorithm \ref{alg_1} a second time. Suppose a scenario in which just one field is removed without its replacement by another field from a tour where all BPs are supplied the exact minimum demand of biomass. Then, it would become infeasible to sufficiently supply one of the BPs for that tour. Thus, no feasible improvement solution could ever be found eventhough it may exist and realised by a simple field exchange. Strictly, a repetition of Step 8 would guarantee feasibility only in expectation  due to the random sampling involved and also only under the assumption that each field produces the same amount of biomass. Otherwise, for example, two smaller fields may be required to replace one larger field and so forth. In simulation experiments the solution with repeated Step 8 was tested. However, it was found that for the given large number of fields there were no symmetry issues and only solve times were prolonged.

The update of $x$ in Step 6 involving the exploration heuristic is implemented outside the GPU on the host. This is done to avoid having to send both $x_\text{old}$ and $x^\star$ to the GPU.

Employment of a GPU permits to conduct Steps 8 of Algorithm \ref{alg_1} in parallel on multiple workers. An implementation detail is discussed. By identical random process control on both GPU and host it is possible to broadcast only the current random seed and $x$ and recover only $C_i$ from all workers $i=1,\dots,n$. Thus, it is not necessary to return all $x_i,~\forall i=1,\dots,n$. The optimal $x_{i^\star}$ can then be reconstructed directly on the host.

Algorithm \ref{alg_1} is terminated after a maximum number of iterations, $N_\text{iters}$. An alternative early termination based on the length of non-decreasing plateaus in the cost function is avoided since it is difficult to tune and would add at least one additional hyperparameter.

In fact, an important benefit of Algorithm \ref{alg_1} is its absence of many hyperparameters. The number of workers $n$ is prescribed by the GPU. Thus, the only two hyperparameters for Algorithm \ref{alg_1} are $\epsilon\in[0,1]$ and $N_\text{iters}>0$. The latter can be selected according to some desired maximum solve time (wall clock time). The former can easily be gridded due to its bounded range.

Algorithm \ref{alg_1} is scalable in that it is inherently parallelisable. Furthermore, there are no steps that are memory-intensive. In addition, there are no steps that involve any special linear algebra tools or other libraries.

Regarding convergence analysis, Algorithm \ref{alg_1} applies iterative local search leveraging random sampling. When leveraging parallel GPU-computations the solution can be interpreted as stochastic global search. Global search is not conducted explicitly on every GPU-worker. However, since GPU-workers are stochastically working together in parallel, global search is conducted implicitly. By non-negativity of path lengths in the cost function and deterministic constraints a global optimal solution exists. All optimisation tasks according to Fig. \ref{Fig_problemVisualisation} are addressed since accumulated total distance is clearly affected by (i) assignments of BPs to tours, (ii) their ordering within each tour, (iii) assignment of fields to BPs, (iv) their ordering for each assignment, (v) assignment of $N_\text{HU}(g),\forall g=1,\dots,N_T$, and (vi) assignment of $N_\text{SU}(g),\forall g=1,\dots,N_T$. Thus, in expectation  Algorithm \ref{alg_1} is capable of finding the global optimal solution.

Finally as a remark, note that the problem of coupling crop assignment and routing for harvest planning discussed in \cite{plessen2019coupling} can be reformulated as a reduced special case of the problem discussed here. Thus, the proposed solution methodology can in general be transfered to related logistics optimisation tasks.

\newlength\figureheight
\newlength\figurewidth
\setlength\figureheight{8.7cm}
\setlength\figurewidth{8.7cm}
\begin{figure}
\centering
\vspace{0.3cm}
\input{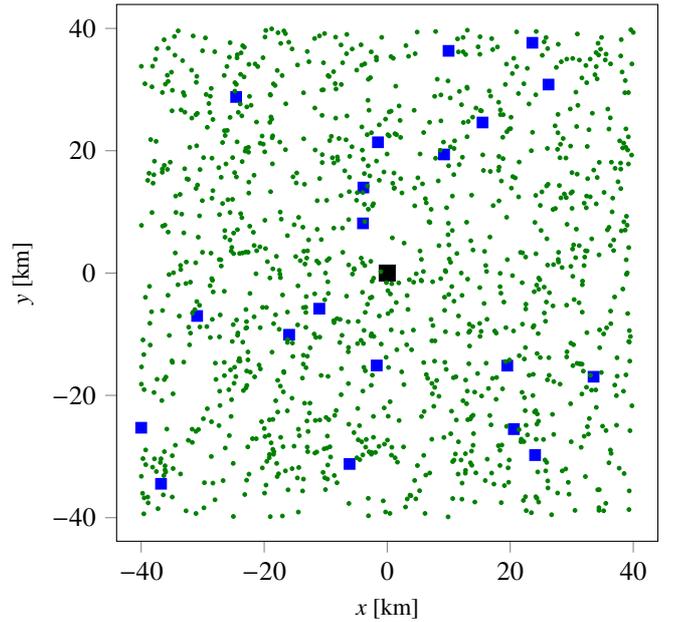}
\caption{Location data set for Experiment 1. Illustration of the distribution of the centroids of 1200 fields (green), 20 biogas plants (blue) and the headquarter (black) within an area of 80km$\times$80km. Field sizes vary between 3 to 7ha. See \S \ref{subsec_ExperimentalSetup} for description of the full experimental setup.}
\label{fig_fig2DAllVertices_Restart9}
\end{figure}

\subsection{Variations and important system parameters\label{subsec_problSoln_variations}}

Different problem variations are discussed. First, consider the important practical scenario in which the assignment of fields to BPs is \emph{fixed}. This occurs if operators of BPs want to draw their supply from very specific fields that they potentially own or have leasing contracts with. Then, the optimisation potential is only with the order in which fields for that BP are harvested to minimise accumulated inter-field distances. In such a scenario, any field removed from $x$ as part of Step 8 in Algorithm \ref{alg_1} must be reinserted into the \emph{same} group of fields assigned to that given BP. This is the only change. The general structure of Algorithm \ref{alg_1} is otherwise not altered.

Second, analogously to the previous scenario one may consider use cases in which the assignment of sets of BPs to a specific tour is fixed, teams of HUs and SUs are fixed, and so forth.

Third, another variation involves the scenario in which in-field operating SUs transfer their load to out-of-field shuttling larger SUs such as trucks that wait along the field boundary. This scenario is mainly relevant if distances between fields and BPs are larger. In the most general case, i.e., if the number of such trucks exceeds the number of tours a new optimisation layer needs to be added for the assignment of these trucks. The proposed optimisation criterion would again be based on a waiting time similar to \eqref{eq_def_tHUwaufbl}, whereby filling times would become a function of the number of the in-field operating SUs and, in general, also a function of the field shape and the in-field path plans driven by HUs and SUs (\cite{plessen2018partial}, \cite{plessen2019optimal}). This scenario is out of the scope of this paper and also not considered for the experimental setup in the next section.

While some parameters such as the number of HUs and SUs or their loading capacities are easy to quantify, there are other parameters that are affected to a larger extent by stochasticity such as seasonal crop yields or soil moisture. These parameters typically are difficult to predict at the time of harvest planning. Due to their importance and for an extending sensitivity analysis in practice they are here mentioned explicitly: $v_\text{HU}^\text{area}$, $t_\text{SU}^\text{load}$ and $c_\text{biom}^\text{conv}$. These are the in-field working rate of HUs, the in-field fill time required by SUs, and the conversion parameter mapping field size to biomass, respectively.

\section{Results and discussion\label{sec_IllustrativeEx}}

\setlength\figureheight{5.7cm}
\setlength\figurewidth{6.9cm}
\begin{figure*}
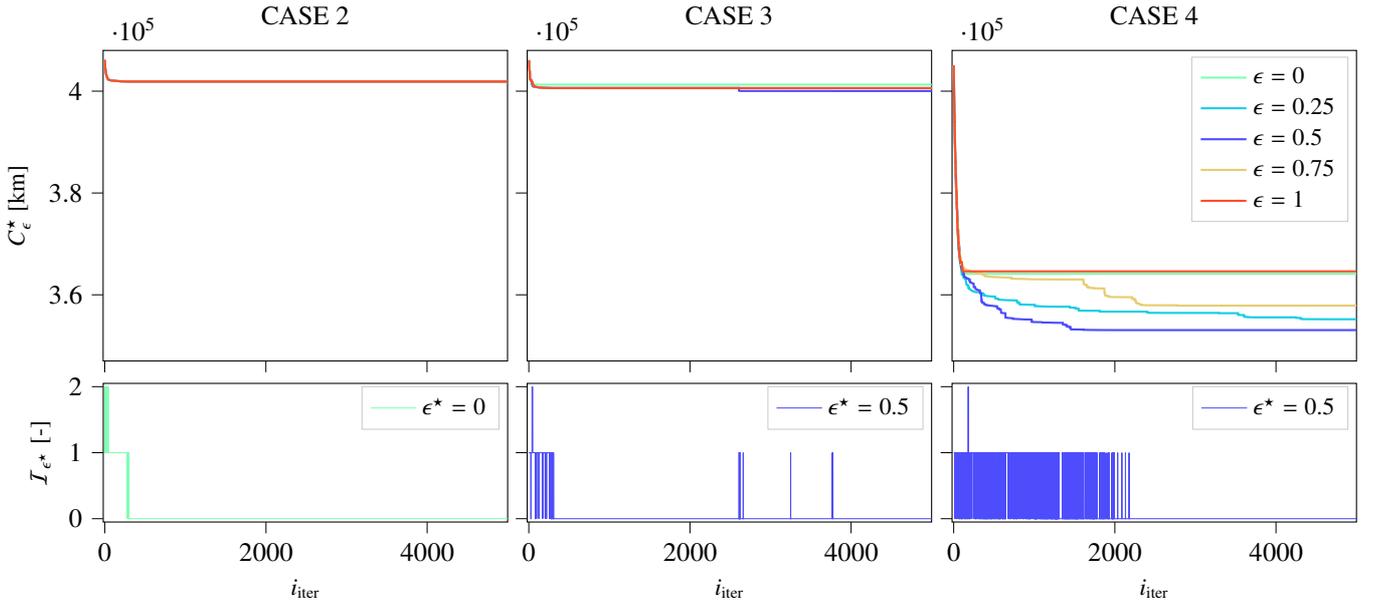

\vspace{-0.2cm}
\begin{tabular}{p{0.3\textwidth} p{0.3\textwidth} p{0.3\textwidth}}
\vspace{0cm}\input{Restart9_Case1.tex} & 
\vspace{0cm}~~~~~~\input{Restart9_Case2.tex} & 
\vspace{0cm}~\input{Restart9_Case3.tex}
\end{tabular}
\caption{Convergence results for Experiment 1 and Cases 2 to 4. Top subplots illustrate the optimal cost $C_\epsilon^\star$ as a function of optimisation iterations, $i_\text{iter}$, for five different $\epsilon$-levels. The legend, which is uniformly valid for all cases, is plotted for clarity only on the right. Bottom subplots illustrate indicator $\mathcal{I}_{\epsilon^\star}$ as a function of optimisation iterations for the cost-minimising best $\epsilon$-level. $\mathcal{I}_{\epsilon^\star}(i_\text{iter})=1$ and $\mathcal{I}_{\epsilon^\star}(i_\text{iter})=2$ imply that $C_\epsilon^\star(i_\text{iter})$ could be decreased by improved reassignment of a field or a biogas plant, respectively. $\mathcal{I}_{\epsilon^\star}(i_\text{iter})=0$ implies that the optimal cost could not be further reduced at iteration $i_\text{iter}$.}
\label{fig_Restart9_Case2to4}
\end{figure*}

\setlength\figureheight{5.45cm}
\setlength\figurewidth{5.45cm}
\begin{figure*}
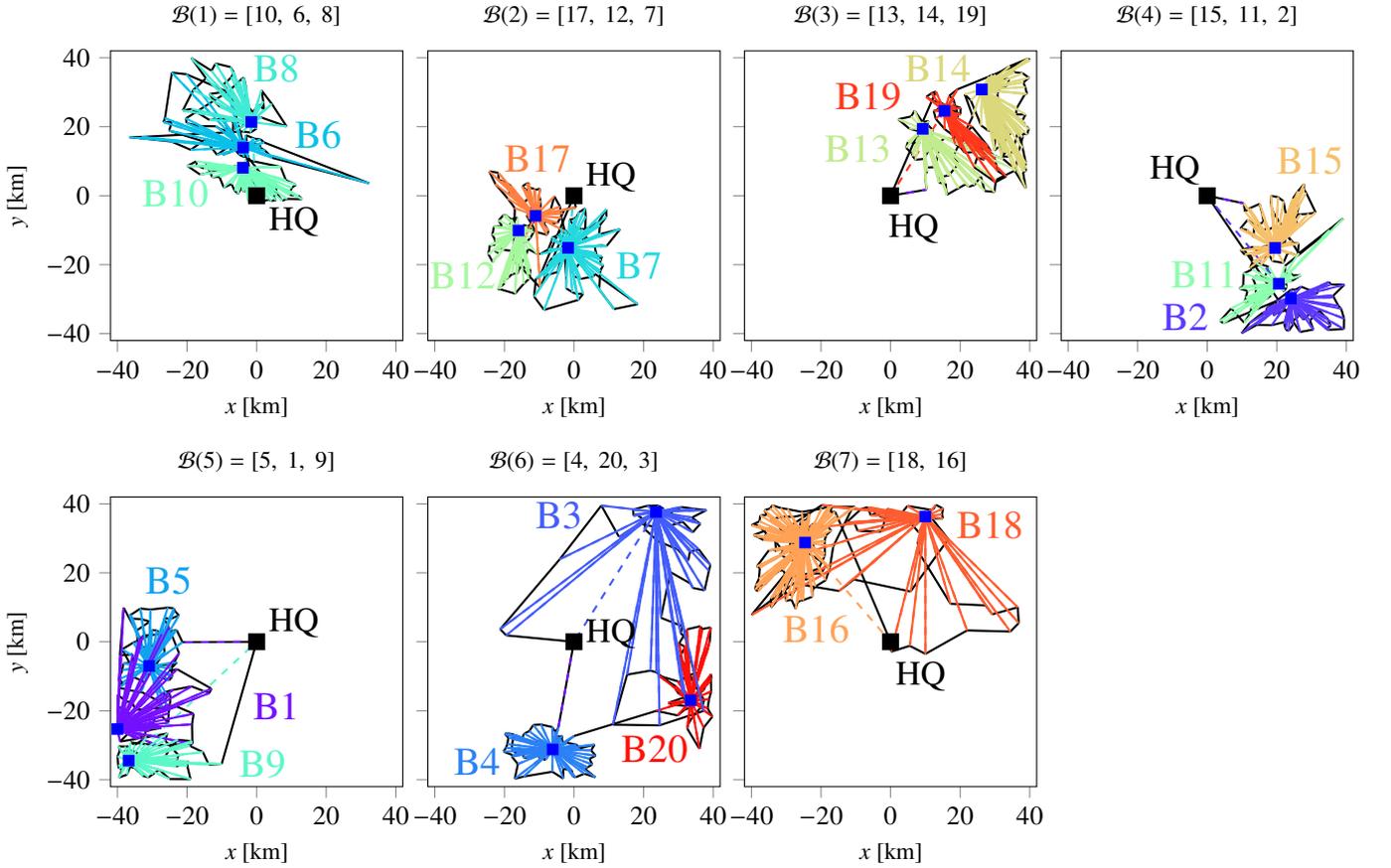

\begin{tabular}{p{0.22\textwidth} p{0.22\textwidth} p{0.22\textwidth} p{0.22\textwidth}}
\vspace{0cm} \input{Plot2D_Restart9_Case3_Grp0.tex} & 
\vspace{0cm} ~~~~~~~~~\input{Plot2D_Restart9_Case3_Grp1.tex} &
\vspace{0cm} ~~~~~\input{Plot2D_Restart9_Case3_Grp2.tex} & 
\vspace{0cm} ~\input{Plot2D_Restart9_Case3_Grp3.tex}\\[-0.2cm]
\vspace{0cm} \input{Plot2D_Restart9_Case3_Grp4.tex} & 
\vspace{0cm} ~~~~~~~~~\input{Plot2D_Restart9_Case3_Grp5.tex} &
\vspace{0cm} ~~~~~\input{Plot2D_Restart9_Case3_Grp6.tex} & \\
\end{tabular}
\caption{Routing results for Experiment 1 and Case 4. For each plot, the solid black line indicates the HU-tour and $\mathcal{B}(g)$ indicates the BPs to be served and their ordering for each tour. Dashed lines indicate the last transition of SUs back to HQ. For a summary of more details see Table \ref{tab_1detailedExperiment}. The initial location  data set is visualised in Fig. \ref{fig_fig2DAllVertices_Restart9}.}
\label{fig_7plots}
\end{figure*}

\begin{table}
\vspace{0.3cm}
\centering
\begin{small}
\bgroup
\def\arraystretch{1.15}
\begin{tabular}{llllll}
\hline
\rowcolor[gray]{0.8} & & \multicolumn{4}{c}{CASE}\\
\rowcolor[gray]{0.9} & Unit & 1 & 2 & 3 & 4 \\\hline
$\epsilon^\star$ & - & - & 0 & 0.5 & 0.5\\
\cline{3-6}
$N_T$ & - & 7 & 7 & 7 & 7 \\
\cline{3-6}
& - & 2 & 2 & 3 & 1 \\
& - & 2 & 2 & 2 & 2 \\
& - & 2 & 2 & 2 & 3 \\
$\mathcal{N}_\text{SU}^\text{small}$& - & 2 & 2 & 1 & 1 \\
& - & 2 & 2 & 2 & 2 \\
& - & 2 & 2 & 2 & 3 \\
& - & 2 & 2 & 2 & 2 \\
\cline{3-6}
& - & 4 & 4 & 4 & 4 \\
& - & 4 & 4 & 3 & 3 \\
& - & 4 & 4 & 4 & 4 \\
$\mathcal{N}_\text{SU}^\text{large}$& - & 4 & 4 & 4 & 4 \\
& - & 4 & 4 & 4 & 4 \\
& - & 4 & 4 & 4 & 4 \\
& - & 4 & 4 & 5 & 5 \\
\cline{3-6}
 & h & 356.2 & 519.2 & 240.4 & 315.6 \\
 & h & 352.2 & 109.3 & 351.4 & 364.5  \\
 & h & 363.5 & 364.5 & 375.5 & 355.8 \\
$\mathcal{T}_\text{HU}^\text{compl}$  & h & 397.2 & 147.5 & 392.8 & 363.1 \\
 & h & 378.9 & 490.1 & 373.6 & 363.3 \\
 & h & 373.3 & 743.9 & 377.1 & 364.3 \\
 & h & 292.6 & 124.0 & 393.9 & 361.8 \\
\cline{3-6}
 & h & 0.07 & 0.02 & -0.04 & 0.01 \\
 & h & -0.13 & -0.14 & 0.01 & -0.04 \\
 & h & 0.03 & 0.2 & 0.04 & -0.10 \\
$\mathcal{T}_\text{HU}^\text{wait}$  & h & -0.09 & -0.10 & 0.00 & -0.05 \\
 & h & 0.08 & 0.03 & 0.07 & -0.04 \\
 & h & 0.06 & -0.03 & 0.05 & -0.09 \\
 & h & 0.23 & 0.12 & 0.03 & -0.07 \\
\cline{3-6}
 & km & 61725 & 83804 & 41934 & 40523 \\
 & km & 43229 & 13083 & 43909 & 41334 \\
 & km & 52944 & 74850 & 61912 & 58371 \\
$\mathcal{C}$  & km& 59911 & 19943 & 49902 & 41780 \\
 & km & 65331 & 79742 & 64040 & 53748 \\
 & km & 61707 & 107866 & 61889 & 57322 \\
 & km & 62082 & 22608 & 76442 & 59991  \\
\cline{3-6}  
$T^\text{solve}$ & s & 0.0049 & 47.5 & 71.5 & 76.9 \\
\cline{3-6}
$C^\star$  & km & 406929 & 401896 & 400028 & 353070 \\
\cline{3-6}
$\Delta C^\star$ & km & - & -5033 & -6901 & \textbf{-53859} \\
$\Delta C^\star$ & $\%$ & - & -1.24 & -1.70 & \textbf{-13.24} \\
\hline
\end{tabular}
\egroup
\end{small}
\caption{Summary of detailed results for one of the 10 experiments. The solutions of Cases 1 to 4 are compared.} 
\label{tab_1detailedExperiment}
\end{table}

\subsection{Experimental setup\label{subsec_ExperimentalSetup}}

The results of 10 simulation experiments are summarised. The task of a contractor about planning logistics and implementing harvest of $N_F=1200$ corn fields for the supply of biomass to $N_B=20$ biogas plants is simulated. System parameters are summarised in Table \ref{tab_problemData}. There is one class of HUs and two classes of SUs, which are for ease of notation in the following abbreviated by $N_\text{SU}^\text{small}(g)$ and $N_\text{SU}^\text{large}(g)$ for tours $g=1,\dots,N_T$. Field sizes are simulated with $h_f=3+4u_f$, measured in hectares (ha), and for uniform random variables $u_f\in[0,1],~\forall f\in\mathcal{F}$. Employing a local practical rule of thumb of one BP requiring a supply of about 250ha of corn fields, the minimum demand of each BP is in simulations also measured in ha and set as $d_b^\text{min}=250$ha, $\forall b\in\mathcal{B}$. Then, the required ``supply'' of each field can conveniently be expressed equivalently as its field size, i.e., $\hat{s}_f=h_f,\forall f\in\mathcal{F}$. In addition, a coverage area of 80km$\times$80km is assumed in which all fields and biogas plants are uniformly distributed. See Fig. \ref{fig_fig2DAllVertices_Restart9} for an example. The headquarter from which all mobile machinery initially starts is simulated to be at the origin. In order to evaluate proposed methods stochastically over 10 different simulation experiments and also for simplicity of generating these scenarios, the edge cost for the connection of any two vertices of the transition graph is calculated as the path length of straight line segments connecting any two vertices. In real-world employment this simplification is to be replaced by nonlinear paths representing the actual road network and involving shortest path computations over the path network in order to compute all edge costs. 

\subsection{Comparison of four solution setups: Case 1 to 4\label{subsec_4Setups}}

Four different solution methods (Case 1 to 4) are compared:
\begin{enumerate}
\item The initialisation method only; see Steps 1-4 of Algorithm \ref{alg_1}. It produces a specific assignment of fields to BPs and enforces a fixed $N_T=N_\text{HU}^\text{total}$ such that $N_\text{HU}(g)=1,~\forall g=1,\dots,N_T$, and a fixed $N_\text{SU}^\text{small}(g)=2$ and $N_\text{SU}^\text{large}(g)=4,~\forall g=1,\dots,N_T$. This method represents the initialisation step of Algorithm \ref{alg_1} and can be interpreted as a concatenation of nearest neighbour methods. This case serves as the baseline reference and replicates a deterministic sequential solution procedure a human scheduler may tacitly conduct.
\item The assignment of fields to BPs generated from Case 1 is kept \emph{fixed}. Likewise, $N_\text{HU}(g)=1$, $N_\text{SU}^\text{small}(g)=2$ and $N_\text{SU}^\text{large}(g)=4,~\forall g=1,\dots,N_T$ remain fixed. Optimisation iterations are permitted to improve (i) the assignment of BPs (and all of their assigned fields) to different tours, and (ii) the ordering of fields (i.e., the harvest sequence) for each BP-assignment. 
\item The assignment of fields to BPs generated from Case 1 is kept \emph{fixed}. However, optimisation iterations according to Steps 5-16 of Algorithm \ref{alg_1} are conducted to improve (i) $N_\text{HU}(g)$, $N_\text{SU}^\text{small}(g)$ and $N_\text{SU}^\text{large}(g),~\forall g=1,\dots,N_T$ , (ii) the assignment of BPs (and all of their assigned fields) to different tours, and (iii) the ordering of fields for each BP-assignment. 
\item The assignment of fields to BPs is now \emph{freely} optimised as part of Steps 5-16 of Algorithm \ref{alg_1}. This adds to the optimisation of (i) $N_\text{HU}(g)$, $N_\text{SU}^\text{small}(g)$ and $N_\text{SU}^\text{large}(g),~\forall g=1,\dots,N_T$ , (ii) the assignment of BPs (and all of their assigned fields) to different tours, and (iii) the ordering of fields for each BP-assignment. 
\end{enumerate}

\subsection{Results of 10 simulation experiments}

The results for one of the 10 simulation experiments are illustrated in detail in Table \ref{tab_1detailedExperiment}, Fig. \ref{fig_Restart9_Case2to4} and \ref{fig_7plots}. The results over all 10 simulation experiments are quantitatively summarised in Table \ref{tab_avgOverAllExperiments}. Note that Algorithm \ref{alg_1} involves only two hyperparameters. For all experiments and solution setups it is set $N_\text{iters}=5000$ and it is uniformly gridded for $\epsilon\in\{0,0.25,\dots,1\}$ to analyse the effect of the exploration heuristic, before the lowest cost solution and corrresponding $\epsilon^\star$ is returned.

A preliminary comment is made. The scope of this section is not to benchmark different solution methods and their solver speeds (e.g., different hybrid genetic algorithms, ant colony optimisation, and so forth). Rather, the scope of this section is to stress structural observations made which are most relevant in practice for contractors who manage logistics with service of many fields and BPs owned by multiple different parties. 

The following observations are made. First, $N_T=7$ in Table \ref{tab_1detailedExperiment} and \ref{tab_avgOverAllExperiments} implies $N_\text{HU}(g)=1,~\forall g=1,\dots,N_T$, and thus $N_T=N_\text{HU}^\text{total}$. While this was expected for Case 1 and 2 by enforcement, interestingly it also resulted for Case 3 and 4 throughout all 10 simulation experiments, eventhough Algorithm \ref{alg_1} is explicitly designed to also permit $N_T<N_\text{HU}^\text{total}$. It implies that every available HU forms exactly one harvester tour and there are never more than one HUs operating together on any tour. The reason for this result is explained by the optimisation criterion designed to minimise not only accumulated path length but also the maximum completion time, $\mathcal{T}_\text{HU}^\text{compl}(g)$, over all tours $g=1,\dots,N_T$. When dropping the latter criterion and just optimising accumulated path length, indeed $N_T<N_\text{HU}^\text{total}$ could be observed. However, then in practice unacceptably unbalanced completion times, $\mathcal{T}_\text{HU}^\text{compl}$, resulted such as, e.g., 800h for some groups and 150h for others. The takeaway message here is that for the given parameters setup the distribution of one HU per tour contributed importantly to achieve approximately uniform completion times over all harvesters tours, with particularly balanced results for Case 4.

Second and in contrast to HUs, subtle differences in the assignment of both small and large SUs to the different groups could be observed for Case 3 and 4 throughout all experiments. See the listings $\mathcal{N}_\text{SU}^\text{small}$ and $\mathcal{N}_\text{SU}^\text{large}$ in Table \ref{tab_1detailedExperiment}. Of course, by construction this does not hold for Case 1 and 2 where the assignments are fixed with $N_\text{HU}(g)=1$, $N_\text{SU}^\text{small}(g)=2$ and $N_\text{SU}^\text{large}(g)=4,~\forall g=1,...,N_T$.

Third, typical convergence results are demonstrated in Fig. \ref{fig_Restart9_Case2to4}. For all Cases 2 to 4 the major reduction in cost $C^\star$ is observed for initial iterations $i_\text{iter}$, before costs typically quickly plateaued. Differences in final cost levels dependent on hyperparameter $\epsilon$ result in particular for Case 4. Interestingly, most of the cost reductions are obtained by field exchanges, i.e., for $\mathcal{I}_{\epsilon^\star}(i_\text{iter})=1$. One explanation is that $N_F \gg N_B$, which implies that fields are drawn much more frequently as candidates for reassignments in comparison to BPs. However, these results also imply that the initialisation method discussed in \S \ref{subsec_mainAlg} for the initial assignment of BPs to groups is quite good. Nevertheless, for all Cases 2 to 4 also at least 1 BP- exchange resulted. Fig. \ref{fig_Restart9_Case2to4} clearly visualises the significant improvement potential of Case 4. The attainable cost level is significantly lowered in comparison to Case 2 and 3, see Fig. \ref{fig_Restart9_Case2to4}. This is enabled by permitting an unconstrained assignment of any field to any BP.

Fourth, typical routing results are demonstrated in Fig. \ref{fig_7plots}. Interestingly, for the given data clustering-like solutions with closely located BPs result for tours $g=1,\dots,5$, whereas residuals are partitioned as shown in the plots for $\mathcal{B}(6)$ and $\mathcal{B}(7)$. As the related Table \ref{tab_1detailedExperiment} shows corresponding path lengths for $\mathcal{C}(6)$ and $\mathcal{C}(7)$ are among the longest. However and crucially, completion times $\mathcal{T}_\text{HU}^\text{compl}(g),~\forall g=1,\dots,N_T$, are still approximately uniform over all tours, see Table \ref{tab_1detailedExperiment}.

Fifth, it is further elaborated on different $\mathcal{T}_\text{HU}^\text{compl}$ observed for Cases 1 to 4. In order to account for all 10 experiments the average quantifiers, $\bar{\mathcal{T}}_\text{HU,avg}^\text{compl}=\frac{1}{10}\sum_{e=1}^{10} \frac{1}{7}\sum_{g=1}^7  \mathcal{T}_\text{HU}^\text{compl}(g)$ and $\bar{\mathcal{T}}_\text{HU,worst}^\text{compl}=\frac{1}{10}\sum_{e=1}^{10} \max_{g\in\{1,\dots,7\}} \mathcal{T}_\text{HU}^\text{compl}(g)$, are defined. As Table \ref{tab_avgOverAllExperiments} demonstrates, the worst results are obtained for Case 2. This was expected since here the assignment of HUs and SUs is fixed, and the optimisation criterion is thus just about accumulated total path length but not accounting for balanced completion times. The best results are clearly obtained for Case 4 with $\bar{\mathcal{T}}_\text{HU,worst}^\text{compl}=371.2$h. In addition, Case 4 also has the lowest $\bar{\mathcal{T}}_\text{HU,avg}^\text{compl}=358.1$h.

\begin{table}
\vspace{0.3cm}
\centering
\begin{small}
\bgroup
\def\arraystretch{1.15}
\begin{tabular}{llllll}
\hline
\rowcolor[gray]{0.8} & & \multicolumn{4}{c}{CASE}\\
\rowcolor[gray]{0.9} & Unit & 1 & 2 & 3 & 4  \\\hline
$\bar{\epsilon}^\star$ & - & - & 0.0 & 0.4 & 0.4 \\
\cline{3-6}
$\bar{N}_T$ & - & 7 & 7 & 7 & 7 \\
\cline{3-6}
$\bar{\mathcal{T}}_\text{HU,avg}^\text{compl}$ & h & 360.9 & 359.8 & 360.2 &358.1 \\
$\bar{\mathcal{T}}_\text{HU,worst}^\text{compl}$ & h & 411.3 & 577.5 & 406.2 & 371.2 \\
\cline{3-6}
$\bar{\mathcal{T}}_\text{HU,avg}^\text{wait}$ & h & 0.12 & 0.08 & 0.09 & 0.01 \\
$\bar{\mathcal{T}}_\text{HU,worst}^\text{wait}$ & h & 0.42 & 0.37 & 0.14 & 0.06 \\
\cline{3-6}
$\bar{T}^\text{solve}$ &  s & 0.0050 & 47.77 & 82.33 & 81.89 \\
\cline{3-6}
$\bar{C}^\star$ & km & 461048.6 & 445043.6 & 446551.3 & 394679.9\\
\cline{3-6}
$\Delta \bar{C}^\star$ & km & - & -16005.0 & -14497.3 & \textbf{-66368.7} \\
$\Delta \bar{C}^\star$ & $\%$ & - & -3.4 & -3.1 & \textbf{-14.4} \\
\hline
\end{tabular}
\egroup
\end{small}
\caption{Summary of results averaged over all 10 simulations experiments. The solutions of Cases 1 to 4 are compared.} 
\label{tab_avgOverAllExperiments}
\end{table}

Sixth, different $\mathcal{T}_\text{HU}^\text{wait}$ for Case 1 to 4 are compared. Similarly to before, $\bar{\mathcal{T}}_\text{HU,avg}^\text{wait}$ and $\bar{\mathcal{T}}_\text{HU,worst}^\text{wait}$, are defined. Table \ref{tab_avgOverAllExperiments} summarises results. Again, Case 4 clearly outperforms the other methods. Interestingly, it achieves $\bar{\mathcal{T}}_\text{HU,avg}^\text{wait}=0.01$h. This implies that on average even for Case 4 there is a tiny delay for HUs operations. This implies that a simulated coverage area of 80km$\times$80km as illustrated in Fig. \ref{fig_fig2DAllVertices_Restart9} would in practice be too large or at least critical for the given machinery and parameters setup according to Table \ref{tab_problemData} in order to operate fully delay-free. For the 10 simulation experiments a large coverage area of 80km$\times$80km was selected to emphasise this aspect.

Seventh, the average solve time, $\bar{T}^\text{solve}$,  is obviously lowest for Case 1  since the method involves no optimisation iterations and no GPU-interaction. $\bar{T}^\text{solve}$ is comparable for Case 3 and 4, and $-42\%$ smaller for Case 2. This is explained by the fact that Case 2 involves in contrast to Case 3 and 4 no embedded optimisation step with respect to the assignment of HUs and SUs to different groups.

Eigth, on average the best hyperparameter choice for Case 3 and 4 is $\bar{\epsilon}^\star=0.4$. This implies to reinitialise $x$ according to \eqref{eq_xUpd_explorHeuristic} in Step 6 of Algorithm \ref{alg_1} in $40\%$ of all $N_\text{iters}$ iterations with the current best solution at that time (exploitation step).

Nineth, it is observed that the cost reduction for Case 2 and 3 in comparison to the benchmark Case 1 is comparatively small with $-3.4\%$ and $-3.1\%$, respectively. Note further that the at first sight surprising result of an on average slightly lower cost for Case 2 rather than for Case 3 is explained by the fact that Case 3 optimises both accumulated total path length and maximum completion time over all 7 tours. Consequently, the worst-case time, $\bar{\mathcal{T}}_\text{HU,worst}^\text{compl}$, is much lower for Case 3 as Table \ref{tab_avgOverAllExperiments} shows.

Tenth, one might consider adding hard constraints such that any GPU solution candidates are only permitted if waiting times over all tours are \emph{negative} such that there is never any HU waiting. This was tested. It was found that in some of the 10 experiments no feasible solution could then be found. Thus, for the given location distribution of 20 BPs and 1200 fields, the given number of HUs and SUs according to Table \ref{tab_problemData}, and in particular for a large coverage area of 80km$\times$80km, in some scenarios there did not exist a solution without any waiting times. Therefore, instead of hard constraints, HUs-waiting time was added as a soft constraint in form of an embedded local minimisation.

Eleventh and to sum up, Case 4 clearly outperforms the other methods. While also offering the best results for $\bar{\mathcal{T}}_\text{HU,avg}^\text{compl}$ and $\bar{\mathcal{T}}_\text{HU,avg}^\text{wait}$, the accumulated path length savings are uncomparably large with -66368.7km or -14.4$\%$ with respect to the baseline reference. While outperformance was in general expected  since Case 4 implies the least constrained setup, the degree of improvement was unexpected. 

In practice, this result is relevant for contractors for managerial insights and importantly also for environmental considerations. It implies that cleanest biomass production from a logistics point of view can only be achieved if all fields a contractor services can be freely assigned to \emph{any} biogas plant independent of their ownerships. Note that all improvements are obtained without compromising the operation of BPs since supply to all BPs is still guaranteed. Thus, the same service is provided cleaner. Furthermore, path length savings directly translate to fuel savings (Diesel) and thus directly to a reduction of exhaust gas and CO$_2$ emissions.

Finally, it is pointed out that results in Table \ref{tab_1detailedExperiment} and \ref{tab_avgOverAllExperiments}, and in particular for Case 4, are bechmarked with respect to an  \emph{optimised} initialisation technique. It is stressed that in practice the fixed assignment of fields to BPs may be a lot worse, for example, due to contractual agreements between BP operators,  field owners, leasing parties and so forth. Thus, path length saving and the environmental benefit are expected to be in practice (much) higher.

%
\section{Conclusions\label{sec_conclusion}}

This paper contributed to the task of cleaner biomass production via logistics optimisation by (i) proposing methods for problem modeling, and (ii) presenting a corresponding parallelisable solution algorithm suitable for GPU-acceleration. Multiple harvesters tours are to be executed simultaneously by groups of harvesting units (HUs) and support units (SUs) to first harvest biomass from multiple agricultural fields, before supplying the biomass to multiple biogas plants (BPs) via shuttling SUs. There are three interconnected optimisation levels: (i) the assignment of BPs to tours and the ordering of BPs assigned to each tour, (ii) the assignment of fields to BPs and the ordering of fields assigned to each BP, and (iii) determining the number of HUs and SUs assigned to each tour, whereby the HUs and SUs may in general have different working rates and loading capacities. In this paper, the optimisation criterion is not only accumulated total path length but also minimisation of the maximum completion time over all harvesters tours, and further, in an embedded optimisation loop, minimisation of waiting times for HUs by non-uniform assignment of SUs to each harvester tour. Large-scale experiments were conducted  with 7 HUs, 14 small SUs, 28 large SUs, 20 BPs and 1200 corn fields operated over an area of 80km$\times$80km. In 10 stochastic simulation experiments four different setups differing in their constraints setup were compared. 

The main managerial insight found was that for one setup (Case 4) significant path length savings can be achieved when permitting free and unconstrained assignment of any of the given fields a contractor services to any of the given biogas plants independent of their ownerships. In addition, this setup permits to simultaneously attain approximately uniform completion times over all harvesters tours and to minimise waiting times by suitable assignment of SUs to different harvesters tours. This result encourages contractors to work towards collaboration of different BP operators, field owners and leasing parties.

In order to evaluate proposed methods stochastically over 10 different simulation experiments and also for simplicity of generating these scenarios, all location vertices (the central depot, biogas plants and fields) were assumed to be connected by straight line segments. In real-world employment this simplification is to be replaced by nonlinear paths representing the actual road network. 

In this study all BPs are supplied with a single crop (corn). For future work, proposed methodology may be adapted to account for alternative BPs that require a mix of multiple different crops and biomasses such as forest residues.

\bibliographystyle{model5-names} 
\bibliography{mybibfile.bib}
\nocite{*}







\end{document}